\documentclass[pra,twocolumn,aps,showpacs]{revtex4}
\usepackage{graphicx}
\usepackage{epsfig}
\usepackage{bm}
\usepackage{amssymb}
\usepackage{mathrsfs}
\usepackage{amsmath}

\begin{document}
\title{Sound speed of a Bose-Einstein condensate in an optical lattice}
\author{Z. X. Liang}
\affiliation{Institute of Physics, Chinese Academy of Sciences, P.O.
Box 603, Beijing 100080, China} \affiliation{Shenyang National
Laboratory for Materials Science, Institute of Metal Research and
International Centre for Materials Physics, Chinese Academy of
Sciences, Wenhua Road 72, Shenyang 110016, China}
\author{Xi Dong}
\affiliation{Institute of Physics, Chinese Academy of Sciences, P.O.
Box 603, Beijing 100080, China}%
\affiliation{Department of Physics, Tsinghua University, Beijing
100084, China}%
\author{Z. D. Zhang}
\affiliation{Shenyang National Laboratory for Materials Science,
Institute of Metal Research and International Centre for Materials
Physics, Chinese Academy of Sciences, Wenhua Road 72, Shenyang
110016, China}
\author{Biao Wu}
\affiliation{Institute of Physics, Chinese Academy of Sciences, P.O.
Box 603, Beijing 100080, China}
\date{\today}

\begin{abstract}
The speed of sound of a Bose-Einstein condensate in an optical
lattice is studied both analytically and numerically in all three
dimensions. Our investigation shows that the sound speed depends
strongly on the strength of the lattice. In the one-dimensional
case, the speed of sound falls monotonically with increasing lattice
strength. The dependence on lattice strength becomes much richer in
two and three dimensions. In the two-dimensional case, when the
interaction is weak, the sound speed first increases then decreases
as the lattice strength increases. For the three dimensional
lattice, the sound speed can even oscillate with the lattice
strength. These rich behaviors can be understood in terms of
compressibility and effective mass. Our analytical results at the
limit of weak lattices also offer an interesting perspective to the
understanding: they show the lattice component perpendicular to the
sound propagation increases the sound speed while the lattice
components parallel to the propagation decreases the sound speed.
The various dependence of the sound speed on the lattice strength is
the result of this competition.
\end{abstract}

\maketitle

\section{Introduction}
A Bose-Einstein condensate(BEC) in an optical lattice has attracted
great interests recently, both experimentally and
theoretically\cite{MorschRev,Bloch}. The presence of the lattice can
remarkably enrich the behaviors of the system compared to the
uniform case, providing a new fertile ground for exploring a variety
of solid-state effects in BECs, for example, Bloch
oscillations\cite{Anderson,Morsch,Choi} and Landau-Zener
tunneling\cite{Loop1,Zobay,Jona,Loop2} between Bloch bands in an
accelerating optical lattice. Moreover, a BEC in an optical lattice
can be considered as quantum simulators and therefore used for
testing fundamental theoretical concepts\cite{Bloch}. For example,
it can be used to simulate the Bose-Hubbard model and study
experimentally the quantum phase transition between superfluid and
Mott-insulator\cite{Jaskch,Greiner}.

In this article, we launch a systematic study of the speed of sound
for a BEC in an optical lattice in all three dimensions. The speed
of sound is important for two simple reasons: (1) it is a basic
physical parameter that tells how fast the sound propagates in the
system; (2) it is intimately related to superfluidity according to
Landau's theory of superfluid. Because of these, the sound
propagation and its speed was one of the first things that have been
studied by experimentalists on a BEC since its first realization in
1995\cite{BEC1995}. The propagation of sound in a harmonically
trapped condensate without the lattice has already been observed
experimentally\cite{Andrews} and studied
theoretically\cite{Zaremba}. Now there are experimental efforts to
measure the sound speed for a BEC in an optical lattice\cite{Du}.

There have been a great deal of theoretical work done to understand
the sound speed for a BEC in an optical lattice. These studies show
that three parameters strongly affect the speed of sound: the
strength of the optical lattice $v$, the interaction between atoms
$c$, and the lattice dimension $D$ ($D=1,2,3$). In Ref.\cite{Berg},
the phonon excitations of the BECs in one-dimensional optical
lattice ($D=1$) are theoretically investigated by solving the
Bogoliubov equations. Their analytic results for the sound speed in
the weak potential limit predicted that the sound speed decreases
monotonically with increasing the depth of the optical lattice. The
most detailed study of sound propagation in one-dimensional lattice
was done by Pitaevskii and Stringari's group\cite{Kraer,Menotti},
who also found the sound speed is suppressed by the lattice. In
particular, Ref. \cite{Kraer} presents the detailed comparison
between the sound speed obtained by Bogoliubov theory with the one
obtained from the compressibility and the effective mass. Similar
results\cite{Danshita} were also obtained for the Kr\"onig-Penney
potential, a special form of the periodic potential. Furthermore,
Martikainen and Stoof\cite{Martikainen} examined the effect of the
transverse breathing mode on the longitudinal sound propagation for
a BEC in a one-dimensional optical lattice. In particular, they
discuss how the coupling with the transverse breathing mode
influences the sound velocity in an optical lattice. Kr\"amer {\it
et al.}\cite{Kramer} also studied the effect of the transverse
degrees of freedom on the velocity of sound of a BEC in 1D optical
lattice and radially confined by a harmonic trap. A recent paper by
Taylor and Zaremba\cite{Taylor} studied the Bogoliubov excitations
of a BEC in an optical lattice in all spatial dimensions. However,
the formulation in Ref. \cite{Taylor} is in principle; the authors
did not present the concrete results of sound speed in two and
three-dimensional cases(D=2,3). Most interestingly, with numerical
calculations Boers {\it et al.}\cite{Boers} found that the sound
speed of a BEC in a three-dimensional optical lattice achieves a
maximum with increasing lattice depth. Because of the difficulty to
obtain the Bloch states with interaction, the investigation of Boers
{\it et al.} is limited to the low density so that the Bloch wave
function of the free particle can be used as an approximation.

Our investigation here tries to overcome the deficiencies in
previous studies to give a complete picture how the sound speed is
affected by the lattice strength $v$, the interaction between atoms
$c$, and the dimensionality $D$. Analytical approaches are used in
two limiting cases: weak lattice and strong lattice. For weak
lattices, they can be viewed as perturbations. In this case, we
obtain an analytical expression to the second order of the lattice
strength for the sound speed of a BEC in {\it
an arbitrary periodic potential}. We have analyzed this result
for the important case of the periodic potential being an optical lattice.
Our analysis finds a strong dependence of
the sound speed on the lattice dimensions. Especially,
we find  that the lattice
component perpendicular to the sound propagation increases the sound
speed while the lattice components parallel to the propagation
suppresses the sound speed. Since the lattice can only be parallel
to the propagation direction of sound in one dimension ($D=1$), the
sound speed falls monotonically with increasing lattice strength. In
two and three dimensions ($D=2,3$), there are both perpendicular and
parallel components in the lattice and, therefore, there is a
competition. As a result, there is a rich dependence of the sound
speed on lattice strength in the case of $D=2,3$. The sound speed
can first increases then decreases as the lattice strength
increases. We have also tried to understand these results from a
different angle, i.e., in terms of compressibility $\kappa$ and
effective mass $m^*$.  The analytical expression is found for
compressibility $\kappa$ and effective mass $m^*$ for a BEC in
an optical lattice. We find that the effect of the lattice on the
sound speed reflects a competition between the slowly decreasing
compressibility $\kappa$, and the increasing effective mass
$m^*$, with increasing lattice depth.

In the limit of strong lattices, it is reasonable to use the
tight-binding model to describe the BEC in an optical
lattice\cite{Smerzi}. Our analytical results display that the sound
speed always exponentially decreases with increasing the optical
lattice in all dimensions. This $D$-independent behavior of sound
speed can be understood as the competition between the tunneling
strength $J$ between adjacent sites and the interaction $U$ between
the atoms at a lattice site. With increasing lattice depth, $U$
slowly increases while $J$ exponentially decreases, resulting in
monotonically decreasing speed of sound.

Our analytical results are complemented by our numerical study,
where the results are obtained for all ranges of lattice strength.
Our numerical results agree well with our analytical results in both
weak potential and tight-binding limits for the case of weak
interatomic interaction. For the intermediate strength of lattice,
we find that the sound speed even oscillates with the lattice
strength for a three-dimensional optical lattice. We emphasize that
in our numerical calculations the interaction between atoms is taken
into account to compute the Bloch states in all three dimensions. In
Ref.\cite{Boers}, the interaction is neglected in computing Bloch
states for BECs and the Bloch states of free bosons were used as an
approximation.

This paper is organized as follows. In Section II, for the sake of
self-containment and introducing notations, we describe the basic
theoretical framework of our study. It includes the definition of
the sound speed $v_{s}$, compressibility $\kappa$, and effective
mass $m^{*}$. In Section III, we present the analytic results of the
sound speed for a BEC in the optical lattice in both weak potential
limit and tight-banding regime. Section IV contains our numerical
study of the sound speed. The details of our numerical methods are
given here. In Section V, we discuss the possibility of observing
the phenomena presented in this paper within the current
experimental capability. The last section (Sec. VI) contains a
discussion of our results and concluding remarks. Five appendices
are given at the end to show the detailed steps to derive our key
analytical results in the main text.

\section{Basic theory}
\subsection{Mean-field theory of Bose-Einstein condensates}
We focus on the situation that the BEC system can be well described
by the mean-field theory. In this case, the BEC system is governed
by the following grand-canonical Hamiltonian,
\begin{eqnarray}\label{GP}
\mathscr{H}&=&\int d^3\vec{r}\Big\{ \psi
^{*}\left(\vec{r}\right)\Big[ -\frac{1}{2} \nabla
^{2}+V_{latt}\left( \vec {r}\right) \Big] \psi\left(\vec{r}\right)\nonumber\\
&&+\frac{c}{2}\left\vert \psi\left(\vec{r}\right) \right\vert
^{4}-\mu \left\vert\psi\left(\vec{r}\right)\right\vert^2 \Big\}.
\end{eqnarray}
In our case, the external potential is a three-dimensional optical
lattice created by six laser beams that are perpendicular to each
other\cite{MorschRev,Bloch},
\begin{equation}\label{pot}
V_{latt}\left(\vec{r}\right)=v\Big[\cos\left(x\right)
+\cos\left(y\right)+\cos\left(z\right)\Big],
\end{equation}
where $v$ characterized the strength of the optical lattice. In Eq.
(\ref{GP}), all the variables are scaled to be dimensionless by the
system's basic parameters: the atomic mass $m$, the wave number
$k_{L}$ of the laser light, and the average density $n_{0}$. The
chemical potential $ \mu $ and the strength $v$ of the periodic
potential are in the units of $ 4\hbar^2 k_{L}^{2}/m$, the wave
function $\psi $ is in the units of $\sqrt{n_{0}}$, $\vec{r}$ is in
the units of $1/2k_{L}$, and t is the units of $m/4\hbar k_{L}^{2}$.
The nonlinear coefficient $c=\pi n_{0}a_{s}/k_{L}^{2}$, where $
a_{s}> 0$ is the $s$-scattering length.

Sound is a propagation of small density fluctuations inside a
system. To study sound in a BEC, one first need to find out the
ground state of this BEC system, which serves as a media for sound
propagation. The sound speed can then be found by perturbing the
ground state as explained in detail in the next subsection.

The ground state of a BEC in an optical lattice is a Bloch state at
the center of the Brillouin zone. Briefly, the Bloch state is of the
following form
\begin{equation}\label{Bloch}
\psi(\vec{r})=e^{i\vec{k}\cdot\vec{r}}\phi_{\vec{k}}(\vec{r}),
\end{equation}
where $\vec{k}$ is the Bloch wave vector and
$\phi_{\vec{k}}\left(\vec{r}\right)$ is a periodic function with the
same periodicity of the optical lattice. The Bloch wave function
$\phi_{\vec{k}}(\vec{r})$ satisfies the following stationary
Gross-Pitaevskii equation
\begin{equation}
-\frac{1}{2}\left(\nabla+i\vec{k}\right)^2\phi_{\vec{k}}
+c|\phi_{\vec{k}}|^2\phi_{\vec{k}}+V_{latt}\left(\vec{r}\right)\phi_{\vec{k}}
=\mu(\vec{k}) \phi_{\vec{k}},  \label{BE}
\end{equation}
where $\mu(\vec{k})$ is the chemical potential. The energy of the
system in a Bloch state is given by
\begin{eqnarray}
E\left(\vec{k}\right)=\int d^3\vec{r}\Big\{ \phi^{* }_{\vec{k}}\big[
-\frac{ \left(\nabla+i\vec{k}\right)^2}{2}&+&V_{latt}\left( \vec
{r}\right) \big]\phi_{\vec{k}}\nonumber\\ &+&\frac{c}{2}\left\vert
\phi_{\vec{k}}\right\vert^{4}\Big\}. \label{b_eng}
\end{eqnarray}
The set of energies $E(\vec{k})$ then forms a Bloch
band\cite{Wu,Smith}. The Bloch state can be obtained analytically in
certain circumstances\cite{Exact}. In most cases, it has to be
computed numerically\cite{Wu,Seaman}. The numerical method of this
study is described in Section VI. To compute the sound speed, one
may only need the Bloch state $\phi_0$ at $\vec{k}=0$. However, for
the effective mass defined by
\begin{equation}
\label{Emass} \frac{1}{m^*}=\frac{\partial
^{2}E\left(\vec{k}\right)}{\partial k^{2}},
\end{equation}
where $k=|\vec{k}|$, one also has to compute Bloch states in the
vicinity of $\vec{k}=0$. In this article, we also study one- and
two-dimensional cases. The one-dimensional optical lattice is given
by
\begin{equation}
\label{latt1}
V(x)=v\cos(x)\,,
\end{equation}
and the two-dimensional optical lattice is given by
\begin{equation}
\label{latt2}
V(x,y)=v[\cos(x)+\cos(y)]\,.
\end{equation}

\subsection{Definitions of the sound speed}
In Section II.A, the BEC system is regarded as a Hamiltonian
system by the grand canonical Hamiltonian (\ref{GP}); the
corresponding Gross-Pitaevskii equation can be obtained by the
variation of the Hamiltonian, $i\partial \psi/\partial t=\delta
H/\delta \psi^{*}$,
\begin{equation}\label{GPE}
i\frac{\partial \psi}{\partial
t}=-\frac{1}{2}\nabla^2\psi+V\left(\vec{r}\right)\psi
+c|\psi|^2\psi.
\end{equation}
The Bogoliubov equations can be determined from the linear stability
analysis of the GP equation (\ref{GPE}). To explore a small
disturbance $\delta \phi_{\vec{k}}\left(\vec{r},t\right)$ at a Bloch
state $\phi_{\vec{k}}\left(\vec{r}\right)$, we write
\begin{equation}
\psi(\vec{r},t)=e^{i\vec{k}\cdot\vec{r}-i\mu
t}\big[\phi_{\vec{k}}\left(\vec{r}\right)+\delta
\phi_{\vec{k}}\left(\vec{r},t\right)\big],
\end{equation}
where the disturbance can be similarly written as
\begin{equation}\label{Disturb}
\delta\phi_{\vec{k}}\left(\vec{r},t\right)=u_{\vec{k}}
e^{i\left\{\vec{q}\cdot\vec{r}-\epsilon\left(\vec{q}\right)t\right\}}+
v^{*}_{\vec{k}}e^{-i\left\{\vec{q}\cdot\vec{r}-\epsilon\left(\vec{q}\right)t\right\}}.
\end{equation}
Plugging Eq. (\ref{Disturb}) into Eq. (\ref{GPE}) and keeping only
the linear terms, we arrive at the Bogoliubov equations\cite{Wu},
\begin{equation}\label{bog}
\delta_{z}M_{\vec{k}}\left(\vec{q}\right)
\begin{pmatrix}
u_{\vec{k}}\cr v_{\vec{k}}
\end{pmatrix}
=\epsilon (\vec{q})
\begin{pmatrix}
u_{\vec{k}}\cr v_{\vec{k}}
\end{pmatrix},
\end{equation}
with
\begin{eqnarray}\label{Matrix}
M_{\vec{k}}\left(\vec{q}\right) =\left(
\begin{array}{cc}
\mathscr{L} \left(\vec{k}+\vec{q}\right)& c\phi_{\vec{k}}^{2}\\
c\phi_{\vec{k}}^{\ast 2} & \mathscr{L}\left(-\vec{k}+\vec{q}\right)
\end{array}
\right),
\end{eqnarray}
and
\begin{eqnarray}
 \delta_{z} =\left(
\begin{array}{cc}
1& 0 \\
0&-1
\end{array}
\right),
\end{eqnarray}
where $\mathscr{L}\left(\vec{q}\right)$ is defined as
\begin{equation}
\mathscr{L}\left(\vec{q}\right)=-\frac{1}{2}(\nabla+i\vec{q})^2+
V\left(\vec{r}\right)-\mu+2c|\phi_{\vec{k}}|^2.
\end{equation}
Note that $\vec{q}$ represents mode of the small perturbations and
is of nature of Bloch wave vector as the matrix $M$ is periodic.

In general there are two equivalent definitions for the sound speed
in a BEC. As sound can be regarded as a long wavelength response of
a system to a perturbation, the sound speed can be extracted from
the excitation of a BEC. According to the Bogoliubov theory, the
excitation energy $\epsilon \left(\vec{q}\right)$ of the BEC in a
Bloch state at $\vec{k}=0$ can be found by solving the eigenvalue
problem of Eq. (\ref{bog}). In terms of these excitations, the sound
speed of a BEC system can be defined as
\begin{equation}\label{Sound2}
v_{s}=\lim_{q \rightarrow 0}{\frac{\epsilon
\left(\vec{q}\right)}{q}},
\end{equation}
where $q=|\vec{q}|$.

The other definition arises when the BEC system is regarded as a
hydrodynamics system. In this context, the sound speed  in a BEC is
given by the standard expression\cite{Pines,Kraer,Menotti}
\begin{equation} \label{sound}
v_{s}=\sqrt{\frac{1}{\kappa m^*}},
\end{equation}
where $\kappa$ is the compressibility of the BEC system and defined
as
\begin{equation}\label{Comp}
\kappa^{-1}=n_{0}\frac{\partial \mu} {\partial n_{0}},
\end{equation}
where the chemical potential $\mu$ and $n_{0}$ is the averaged
density. For a BEC system with repulsive interatomic interaction,
the optical trapping reduces the compressibility of the system as
the effect of the repulsion is enhanced by squeezing the condensate
in each well. According to the definition of sound speed in Eq.
(\ref{sound}), the sound speed reflects the competition between the
compressibility $\kappa$ and the effective mass $m^*$.

Both definitions are used in our computations and they agree with
each other as expected. The proof of the equivalence of these
definitions can be found in Ref. \cite{Pines}.

\section{Analytical results}
\subsection{Weak potential limit.}
We consider first  an arbitrary periodic potential
$V_{ar}\left(\vec{r}\right)$ with the periodicity of $\vec{R}$,
\begin{equation}
V_{ar}\left(\vec{r}\right)=V_{ar}\left(\vec{r}+\vec{R}\right),
\end{equation}
with
\begin{equation}
\vec{R}=m_{1}\vec{a}_{1}+m_{2}\vec{a}_{2}+m_{3}\vec{a}_{3},
\end{equation}
where $\vec{r}$ is the position vector, $\vec{a}_{1}$,
$\vec{a}_{2}$, and $\vec{a}_{3}$ are the three primitive vectors and
$m_{1}$, $m_{2}$, $m_{3}$ range through all integral values. In the
weak potential limit, the periodic potential
$V_{ar}\left(\vec{r}\right)$ can be regarded as a perturbation. This
allows us to solve  both the Gross-Pitaevskii equation (\ref{BE})
and the Bogoliubov eigenvalue problem (\ref{bog}) perturbatively by
expanding the wave function $\psi$ and chemical potential $\mu$ of
the BEC system in the order of the weak potential,
\begin{eqnarray}\label{pert}
\psi&=&\psi ^{\left( 0\right) }+\psi ^{\left( 1\right) }+\psi
^{\left( 2\right) }+...,\nonumber\\
\mu &=&\mu ^{\left( 0\right) }+\mu ^{\left( 1\right) }+\mu ^{\left(
2\right) }+....,
\end{eqnarray}
where $\psi ^{\left( 0\right) },\mu ^{\left( 0\right) }$ is zeroth
order of the potential strength, $\psi ^{\left( 1\right) },\mu
^{\left( 1\right) }$ first order, etc. We find that the sound
velocity along a given direction indicated by a unit vector
$\hat{r}$ is
\begin{eqnarray}
v_{s}=\sqrt{c}&+&8\sqrt{c}\sum_{\vec{n}\ne0}\Big\{\frac{|\vec{n}|^2}
{\left(4c+|\vec{n}|^2\right)^{3}}\nonumber\\
&-&\frac{\left|\vec{n}\cdot\hat{r}\right|^{2}}
{|\vec{n}|^2\left(4c+|\vec{n}|^2\right)^{2}}\Big\} {\mathscr F_{\vec
n}}^2(V).\label{Gsound}
\end{eqnarray}
In the above, $\mathscr F_{\vec n}(V)$ is the Fourier coefficient of
$V_{ar}\left(\vec{r}\right)$ as defined by
\begin{equation}
\mathscr F_{\vec n}(V)=\frac{1}{\Omega}\int_{cell} d^3\vec{r}
V_{ar}(\vec{r})e^{-i\vec{n}\cdot\vec{r}},
\end{equation}
with
\begin{equation}
\vec{n}=n_{1}\vec{b}_{1}+n_{2}\vec{b}_{2}+n_{3}\vec{b}_{3}
\end{equation}
where $n_{j}$'s are integers and $\vec{b}_{j}$'s are
the set of reciprocal primitive vectors defined by
\begin{equation}
\vec{a}_{i}\cdot\vec{b}_{j}=2\pi\delta_{ij}.
\end{equation}
In the integration, $\Omega$ is the volume of the primitive cell and
the integration is over one primitive cell. The detailed derivation
of Eq. (\ref{Gsound}) can be found in the Appendix C.

% and D, where
% the two definitions of sound speed (Eqs. (\ref{Sound2}) and
%(\ref{sound})) are used, respectively.}

The focus of this article is optical lattices as described in
Eqs.(\ref{pot},\ref{latt1},\ref{latt2}). In this special but
important case, the primitive vectors $\vec{a}_{1}$, $\vec{a}_{2}$,
and $\vec{a}_{3}$ can be chosen along the directions of laser beams,
$\vec{x}$, $\vec{y}$, and $\vec{z}$, respectively. Also we have
$|\vec{a}_{1}|=|\vec{a}_{2}|=|\vec{a}_{3}|=2\pi$. For this case, we
find from Eq.(\ref{Gsound}) that if the sound propagation direction
is along the $x$-axis, the sound speed is (see also Eq. (\ref{C11})
in the Appendix C or Eq. (\ref{A99}) in the Appendix D)
\begin{eqnarray}
v_{s}=\sqrt{c}+\sum_{\vec{n}\ne0}\frac{8\sqrt{c}
\Big[\left(n_2^2+n_3^2\right)|\vec{n}|^2-4cn_1^2\Big]}
{|\vec{n}|^2\left(4c+|\vec{n}|^2\right)^3}{\mathscr F_{\vec
n}}^2(V). \label{GeneralSound}
\end{eqnarray}
The sound speeds along the $y$-axis and $z$-axis can be found
easily with permutation argument and the sound speed along
a general direction is a certain combination of these three
speeds.

When there is no periodic potential $V_{ar}\left(\vec{r}\right)=0$,
the sound speed in Eq. (\ref{GeneralSound}) is reduced into
$\sqrt{c}$, the sound speed for a BEC in free space, as expected. We
also notice that there is no first-order correction to the sound
speed due to the periodic potential. Most importantly,  the
analytical result in Eq. (\ref{GeneralSound})  reveals that the
lattice component perpendicular to the sound propagation (generated
by the laser beams along the $y$ and $z$-axes) increases
the sound speed while the lattice components parallel to the
propagation (generated by the laser beams along the $x$-axis)
decreases the sound speed. As a result of this competition,
the sound speed can either increase or decrease with lattice strength.
This competition between the parallel and perpendicular components
of the optical lattice certainly also applies to a general
periodic potential if one carefully examines Eq.(\ref{Gsound})
and interprets ``parallel'' and ``perpendicular'' in a more
sense.

To illustrate this more clearly, we consider a simple case where the
periodic potential is a 1D optical lattice given by $V_{ex}(\vec{r})=v\cos(y)$. There are
only two non-vanishing Fourier coefficients, i.e. $\mathscr
F_{0,1,0}(V)= \mathscr F_{0,-1,0}(V)=v/2$. Then according to Eq.
(\ref{GeneralSound}), the speeds of sound along the $x$, $y$ and
$z$-axis read, respectively,
\begin{equation}
\left\{
\begin{array}{rcl}
v_{x,z}&=&\sqrt{c}\left( 1+\frac{v^{2}}{2\left(2c+\frac{1}{2}\right)
^{3}}\right),\\
v_{y}&=&\sqrt{c}\left( 1-\frac{2cv^{2}}{\left(2c+\frac{1}{2}\right)
^{3}} \right),
\end{array}
\right.\label{Example}
\end{equation}
which show  that with increasing the strength of the optical lattice
the sound speed along the $y$-axis, parallel to the periodic lattice
falls while the speeds of sound  along both the $x$ and $z$-axes
increase.

Now we study the BEC sound speed in optical lattices in terms of
compressibility and effective mass according to the
second definition of speed of sound, i.e. Eq.(\ref{sound}).
Again we treat the weak optical lattice as a perturbation.
For optical lattices of all dimensions as
described in Eqs.(\ref{pot},\ref{latt1},\ref{latt2}),
we find the chemical potential at $\vec{k}=0$
\begin{equation}
\mu = c-\frac{Dv^{2}}{4\left( 2c+\frac{1}{2}\right) ^{2}},
\end{equation}
and the system energy near  $\vec{k}=0$
\begin{equation}
E\left( k\right)
=\frac{k^{2}}{2}-\frac{v^2}{\left(1+4c\right)^2\left(4k^{2}-1\right)}.
\end{equation}
So, the chemical potential depends on $D$, the dimension of the
lattice, while the energy $E(k)$ does not. The compressibility
$\kappa $ can be calculated from the chemical potential $\mu$ via
Eq.(\ref{Comp}) and it is given by
\begin{equation}
\kappa ^{-1}= c+\frac{Dcv^2}{\left( \frac{1}{2}+2c\right) ^{3}}\,.
\label{compress}
\end{equation}
This shows that the compressibility $\kappa$ tends to decrease with
increasing $v$ as the optical lattice localizes the BECs inside each
well. Moreover, the compressibility $\kappa$ decreases faster with
$v$ for higher dimensional lattices. The effective mass $m^*$ can be
computed from $E(k)$ and it is found that
\begin{equation}
\frac{1}{m^{\ast }}=\left( 1-\frac{2v^{2}}{\left(
\frac{1}{2}+2c\right) ^{2}}\right). \label{mass}
\end{equation}
It is clear that the effective mass always increases with the
lattice strength $v$. This is expected as the increased lattice
strength suppresses the tunneling between neighboring wells thus
increases the effective mass $m^{*}$. Interestingly, in contrast to
the chemical potential $\mu$, the dependence of the effective mass
on $v$ is independent of the lattice dimension $D$. As the speed of
sound is defined as $v_{s}=\sqrt{1/(\kappa m^*)}$, the
compressibility $\kappa$ and the effective mass $m^{*}$ influence
the sound speed in opposite directions.

Plugging both Eqs.(\ref{compress}) and (\ref{mass}) into Eq.
(\ref{sound}), we find an analytical expressions for the sound speed
of a BEC in an optical lattice up to the second order of $v$
\begin{equation}
v_{s}= \sqrt{c}\left( 1+\frac{4(D-1-4c)}{\left(4c+1\right)
^{3}}v^{2}\right)\,. \label{Wsound}
\end{equation}
With simple algebra (see Eq. (\ref{E6}) in Appendix E), one can show
that this expression is consistent with the more general formula in
Eq.(\ref{GeneralSound}).

Eq.(\ref{Wsound}) indicates that in one dimension ($D=1$) the
effective mass $m^*$ always wins over the compressibility $\kappa$,
resulting in decreasing sound speed with the lattice strength.
However, in two or three dimensions ($D=2,3$), the situation is very
different. There exists a critical value of $c$, the interatomic
interaction strength, beyond which the effective mass $m^*$ wins.
Otherwise, the compressibility $\kappa$ has bigger influence on the
speed of sound and the speed of sound increases as the lattice
becomes stronger. The critical values are $c=1/4$ for $D=2$ and
$c=1/2$ for $D=3$.

We have discussed the behavior of the speed of sound in two
different languages: one in terms of perpendicular and parallel
components of the periodic potential with Eq.(\ref{GeneralSound})
and the other in terms of effective mass $m^*$ and the
compressibility $\kappa$. Are these two pictures consistent? The
answer is yes. To see this, we re-write Eq.(\ref{GeneralSound}) as
\begin{eqnarray}
v_{s}&=&\sqrt{c}+8\sqrt{c}\Big\{\sum_{\vec{n}\ne0}\frac{|\vec{n}|^2}
{(4c+|\vec{n}|^2)^{3}}\nonumber\\
&&-\sum_{\vec{n}\ne0}\frac{n_1^2}{|\vec{n}|^2(4c+|\vec{n}|^2)^{2}}
\Big\}{\mathscr F_{\vec n}}^2(V). \label{GeneralSound2}
\end{eqnarray}
By comparing it with Eqs.(\ref{compress}) and (\ref{mass}), it is
apparent that the first term in the curly brackets comes from the
compressibility $\kappa$ while the second term results from the
effective mass $m^*$. This observation gives us the following
picture: on the one hand, all components of the periodic potential
contribute to the compressibility $\kappa$, which leads to the
enhance of the sound speed; on the other hand, only component
parallel to the sound propagation increases the effective mass
$m^*$, which leads to the suppression of the sound speed. Since the
effective mass $m^{*}$ always wins over $\kappa$ along the parallel
direction, we come to the previous understanding: the perpendicular
components increase the sound speed while the parallel one
suppresses it.

\subsection{Tight-binding regime} When the potential wells are
sufficiently deep, the condensate is well localized at each lattice
site and the following tight-binding model may become adequate to
describe the BEC in an optical lattice\cite{Smerzi}
\begin{eqnarray}
\mathscr{H}=-J\sum_{<\vec{n},\vec{n}^\prime>}\left(\psi_{\vec{n}}^{*}
\psi_{\vec{n}^\prime}+\psi_{\vec{n}^\prime}^{*}\psi_{\vec{n}}\right)
+\frac{U}{2}\sum_{\vec{n}}|\psi_{\vec{n}}|^4, \label{three}
\end{eqnarray}
where the first summation is over all pairs of the nearest
neighbors. The tunneling constant $J$, which quantifies the
microscopic tunneling rate between adjacent sites, is given by
\begin{equation}
J =-\frac{1}{(2\pi)^D}\int d^D\vec{r}\left[\frac{1}{2}\left(
\vec{\nabla}\varphi _{\vec{n}}\cdot \vec{\nabla}\varphi
_{\vec{n}+1}\right) +\varphi _{\vec{n}}V\varphi _{\vec{n}+1}\right],
\end{equation}
with $\varphi _{\vec{n}}$ being the wave function localized at site
$\vec{n}$. The on-site interaction as given by
\begin{equation}
U =\frac{c}{(2\pi)^D}\int d^D\vec{r}\varphi _{\vec{n}}^{4}.
\end{equation}
is a measure of the interaction between atoms at one lattice site.
The ground state of this Hamiltonian is a constant wave function
$\psi_{\vec{n}}=1$. Its excitation energy is given by $\epsilon(q)
=2|\sin \left( q \pi\right)|\sqrt{2 J U }$, which yield the sound
speed via Eq.(\ref{Sound2})
\begin{equation}
v_{s}=\lim_{q\rightarrow 0}\frac{\omega \left( q\right) }{q}=\sqrt{
8\pi^2 J U }\,. \label{Tsound}
\end{equation}
This result is consistent with the sound speed definition in terms
of compressibility $\kappa$ and effective mass $m^*$ since
\begin{equation}\label{38}
J=\frac{1}{8\pi^2}\frac{m}{m^*}\,, ~~~~~~U\approx \kappa^{-1}\,.
\end{equation}

In the following, we try to express $J$ and $U$ in terms of $c$ and
$v$. For a state well-localized at each lattice site, we can regard
it as the ground state of the lattice well and describe the
localized state $\varphi_{\vec{n}}$ with the ground state wave
function of a harmonic oscillator. This approximation immediately
leads to an estimate of $U$. We obtain
\begin{equation}
\label{U} U=c\left(4\pi^2 v\right)^{D/4}.
\end{equation}
As $U\approx \kappa^{-1}$, this indicates that the compressibility
in the tight-binding limit is very similar to the weak potential
limit: it depends on the dimensionality of the lattice and decrease
with $v$ in a non-exponential form.

Mathematically, the time-independent Schr\"odinger equation for an
atom in the cosine potential is a Mathieu equation. The theory of
the Mathieu equation allows us to estimate $J$, which is given
by\cite{Zwerger,Abramowitz}
\begin{equation}
\label{J} J=\frac{4}{\sqrt{\pi}} v^{3/4}\exp\left[-8\sqrt{v}\right].
\end{equation}
This result is very different from the weak potential limit: the
effective mass $\frac{m^*}{m}=\frac{1}{8\pi^2J}$ increases
exponentially with $\sqrt{v}$. As a result, $J$ should dominate the
behavior of the speed sound. Combining Eq.(\ref{U}) and
Eq.(\ref{J}), we arrive at
\begin{equation}
v_{s}=2^{5/2}\pi^{3/4}c^{1/2}(4\pi^2v)^{D/8}v^{3/8}\exp
(-4\sqrt{v}),
\end{equation}
which shows the speed of sound decreases monotonically with $v$ in
an exponential form in all three dimensions. The sound speed in the
tight-binding limit has a weak dependence on the dimension $D$ of
the lattice as $D$ only appears in the prefactor of the exponential.

\section{Numerical results}
We have so far studied analytically the sound speed of a BEC in an
optical lattice. In this section, we study the sound speed with numerical
methods. Our numerical method allows us to
find the sound speed for the whole range of lattice strength,
particularly, intermediate lattice strength for which no apparent
analytical approach can be found.

\subsection{Numerical methods}
As discussed in Section II, to compute the sound speed, one
has to first find the ground state of the BEC system
or the Bloch states in the vicinity of $\vec{k}=0$.
To find these states numerically, we expand the Bloch states
in Fourier series
\begin{equation}
\phi_{\vec{k}}(\vec{r})=
\sum_{m,n,l=-N}^{N}a_{m,n,l}e^{i\left(mx+ny+lz\right)},
\end{equation}
where $N$ is the cut-off. We find that $N=5$ is good enough for all
dimensions. The Fourier coefficients $\{a_{m,n,l}\}$ satisfy the
normalization condition
\begin{equation}
\sum_{m,n,l=-N}^{N}|a_{m,n,l}|^2=1.
\end{equation}
Note that the Fourier coefficients $\{a_{m,n,l}\}$ can be chosen as
real. This fact greatly reduces the computation burden.

The Bloch waves can be numerically obtained by varying
$\{a_{m,n,l}\}$ so that the wave function $\phi_{\vec{k}}$ minimizes
the system energy of Eq.(\ref{b_eng}); the accuracy
is checked by substituting the solutions into the Gross- Pitaevskii
equation (\ref{BE}). We use the standard minimization routine of
MATLAB. The accuracy of the numerical solutions can be checked by
substituting the numerical solutions into the time-independent
Gross-Pitaevskii Eq.(\ref{BE}). Once the Bloch states
$\phi_{\vec{k}}(\vec{r})$ have been obtained, we can compute the
sound speed in two different methods. In one method, we calculate
the Bogoliubov excitations $\varepsilon(q)$ of the ground state
$\phi_0(\vec{r})$ and find the sound speed of the BEC through Eq.
(\ref{Sound2}). In the other method, we can calculate the effective
mass $m^*$ and compressibility $\kappa$, respectively, with
Eqs.(\ref{Emass}) and (\ref{Comp}). Then the sound speed can be
computed via Eq. (\ref{sound}). We have calculated the sound speeds
with both methods and the agreement is excellent as expected.

\subsection{Results and discussion}
\begin{figure}[tbh]
\includegraphics[width=8.0cm]{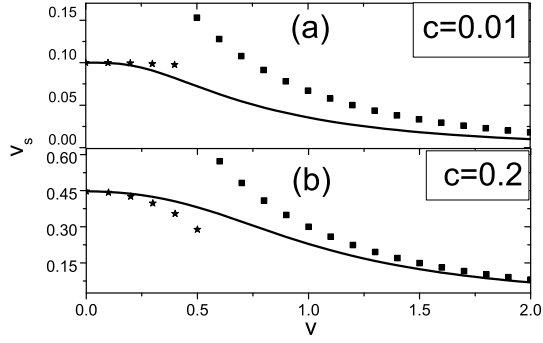}
\caption{Sound speed for a BEC in a 1D optical lattice via the
strength of the optical lattice. (i) numerical result (solid line
curve); (ii) analytical results for weak potentials ($\star$ curve);
(iii) analytical tight-binding results($\blacksquare$ curve).}
\label{fig:sound1D}
\end{figure}

\begin{figure}[tbh]
\includegraphics[width=8.0cm]{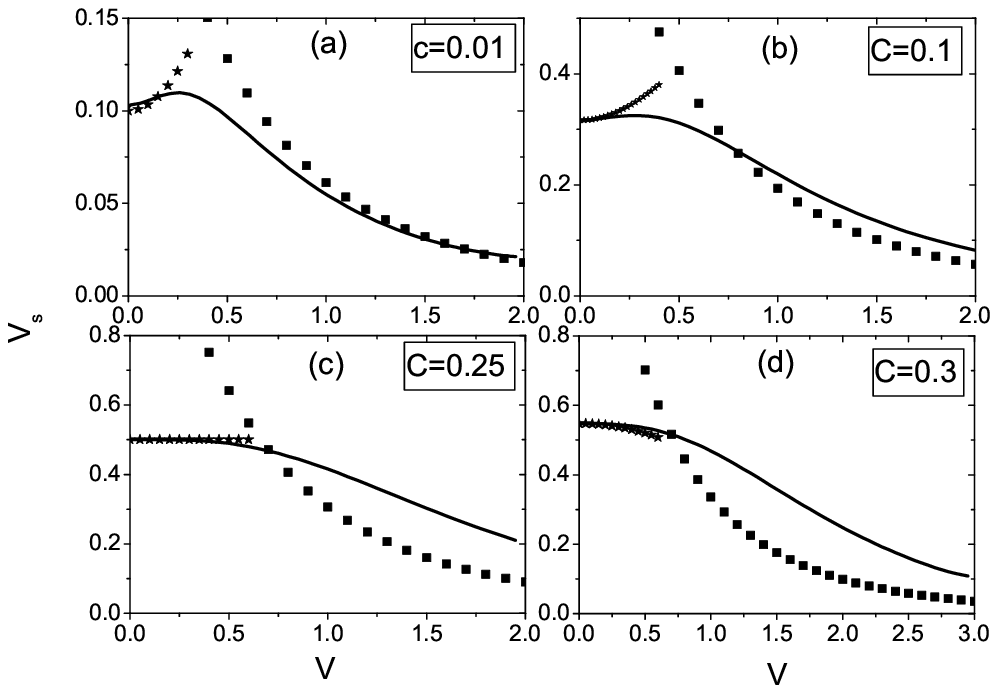}
\caption{Sound speed for a BEC in a 2D optical lattice via the
strength of the optical lattice. (i) numerical result (solid line
curve); (ii) analytical results for weak potentials ($\star$ curve);
(iii) analytical tight-binding results($\blacksquare$ curve). }
\label{fig:sound2D}
\end{figure}

\begin{figure}[tbh]
\includegraphics[width=8.0cm]{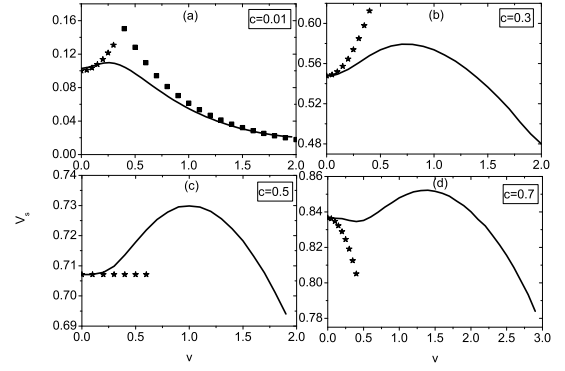}
\caption{Sound speed for a BEC in a 3D optical lattice via the
strength of the optical lattice. (i) numerical result (solid line
curve); (ii)analytical results for weak potentials ($\star$ curve);
(iii) analytical tight-binding results($\blacksquare$ curve).}
\label{fig:sound3D}
\end{figure}

We have computed numerically the sound speeds for all three dimensions
for a wide range of lattice strength $v$ and inter-atomic interaction $c$.
The results are plotted in Figs.\ref{fig:sound1D},\ref{fig:sound2D} \& \ref{fig:sound3D},
respectively. Fig. \ref{fig:sound1D} displays the sound
speed in the one-dimensional case, which falls monotonically with
increasing lattice strength. This is in agreement of previous
studies\cite{Berg,Kraer,Menotti}.
%------------------------------------------------------------------------
%------------------------------------------------------------------------
%-----------------------------------------------------------------------

The results are different in two and three dimensions. As shown in
Figs. \ref{fig:sound2D} and \ref{fig:sound3D}, the relationship
between the sound speed and the lattice strength depends crucially
on the strength of interatomic interaction. In the two-dimensional
case, when the interaction is above the critical value, i.e.
$c>\frac{1}{4}$, the speed of sound also decreases monotonically
with increasing lattice strength (Fig.\ref{fig:sound2D}(d)).
However, when the interaction is weak, i.e. $c<\frac{1}{4}$, as
shown in Fig. \ref{fig:sound2D}(a,b), a sound speed reaches a
maximum at an intermediate strength of optical lattice.
Fig.\ref{fig:sound2D}(c) shows the transition point between the
above two different behaviors, where the sound speed changes almost
does not change with weak lattice potentials.

In three dimensions, the behavior becomes even richer. There exists
a critical value of the interatomic interaction, $c=\frac{1}{2}$.
When the interaction is smaller than this critical value
$c<\frac{1}{2}$, the sound speed first increases then decreases as
the lattice strength increases (Fig. \ref{fig:sound3D}(a,b,c)). This
is similar to the two-dimensional case and was first noticed by
Boers {\it et al.}\cite{Boers}. However, when the interaction is
strong enough, i.e. $c>\frac{1}{2}$, a new pattern is found. As
shown  in Fig. \ref{fig:sound3D}(d), the sound speed can even
oscillate with the lattice strength. According to our numerical
results, the oscillating behavior of the sound speed does not
disappear until the interatomic interaction $c$ reaches 1 ($c=1$).

Our numerical results are compared to our analytical results. As
seen in Figs. \ref{fig:sound1D}, \ref{fig:sound2D} and
\ref{fig:sound3D}, our numerical results agree very well with our
analytical results($\star$ curves) in the regime of weak potentials.
For strong lattices, our numerical results also agree well with the
tight-binding results ($\blacksquare$ curve in Figs.
\ref{fig:sound1D}, \ref{fig:sound2D}(a) and \ref{fig:sound3D}(a))
for weak interactions. However, for strong interaction in two and
three dimensions, there exist discrepancy between the tight-binding
results and our numerical results. This discrepancy is very large
especially for three dimensions as shown in Fig.
\ref{fig:sound3D}(b,c,d). The agreement can only be best described
as  qualitative.

\begin{figure}[tbh]
\includegraphics[width=8.0cm]{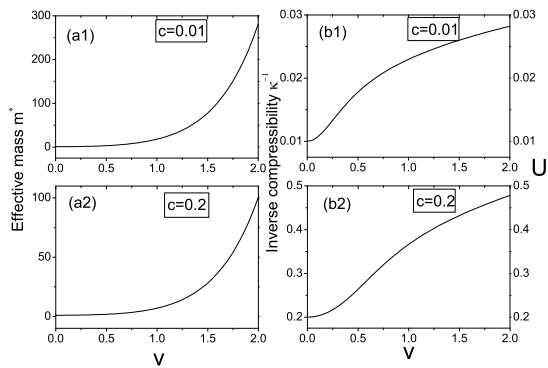}
\caption{Inverse compressibility $\kappa^{-1}$, effective mass $m^*$
and on-site interaction energy $U$ for a BEC in a 1D optical lattice
via the strength of the optical lattice.}\label{fig:CM1D}
\end{figure}

\begin{figure}[tbh]
\includegraphics[width=8.0cm]{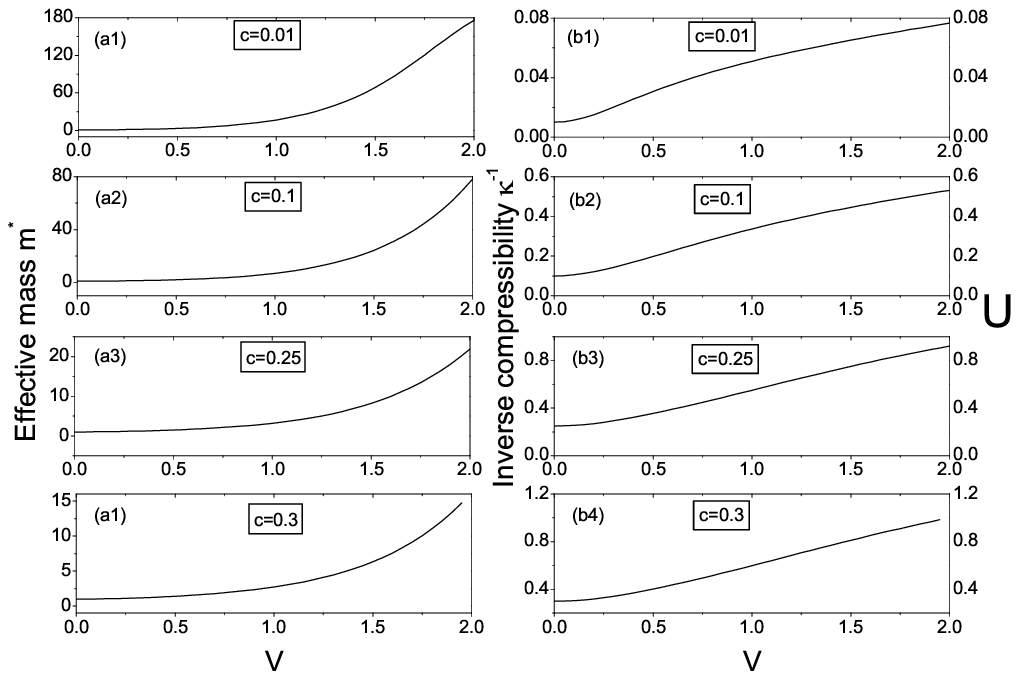}
\caption{Inverse compressibility $\kappa^{-1}$, effective mass $m^*$
and on-site interaction energy $U$ for a BEC in a 2D optical lattice
via the strength of the optical lattice.}\label{fig:CM2D}
\end{figure}

\begin{figure}[tbh]
\includegraphics[width=8.0cm]{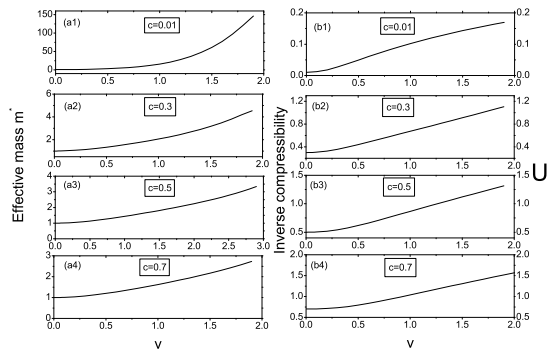}
\caption{Inverse compressibility $\kappa^{-1}$, effective mass $m^*$
and on-site interaction energy $U$ for a BEC in a 3D optical lattice
via the strength of the optical lattice.}\label{fig:CM3D}
\end{figure}

 This big mismatch is nevertheless expected as both
Eq.(\ref{U}) and Eq.(\ref{J}) are derived without the consideration
of interaction between atoms. It can be explained by noticing that
the interaction can strongly modify the ground state wavefunction
localized in each well and greatly enhance the tunneling between the
adjacent wells in two and three dimensions. As in Eq. (\ref{38}),
the tunneling rate $J$ is related to the effective mass $m^*$, which
is very sensible to the behaviour of the wave function in the region
of the barriers. In Figs. \ref{fig:CM1D}(a1)-(a2), Figs.
\ref{fig:CM2D}(a1)-(a4) and Figs. \ref{fig:CM3D}(a1)-(a4), we show
the effect of the interaction $c$ on the effective mass $m^{*}$. The
effect is different in different dimensions. For one dimension, as
shown in Figs. \ref{fig:CM1D}(a1)-(a2), the effective mass $m^{*}$
increases greatly with the lattice strength for both weak and strong
interaction. In 2D, the interaction has much stronger influence on
$m^*$. If we compare Fig.\ref{fig:CM2D}(a1) and
Fig.\ref{fig:CM2D}(a4), the influence is order of magnitude
different. In 3D, the interaction $c$ affect the effective mass
$m^*$ most strongly. As shown in Fig. \ref{fig:CM3D}(a1), for weak
interaction $c=0.01$, the effective mass increases to $\sim 150$ at
$v=2$. In Fig. \ref{fig:CM3D}(a4) for $c=0.7$, the effective mass
$m^*$ is only $\sim 3$ at $v=2$. This order of magnitude difference
explains why there are large disagreement between our numerical
results and the tight-binding model for the speed of sound in
Fig.\ref{fig:sound3D}(b-d).

The dependence of sound speeds on the lattice strength $v$ is
largely expected from our analytical results for the two limiting
cases of weak and strong lattices. We have shown that in the weak
lattice limit, the speed of sound can either increase or decrease
with the lattice strength while in the strong lattice limit the
speed of sound always decreases with increasing lattice strength.
Naively, one would expect that the speed of sound either decreases
monotonically with the lattice strength $v$ or develops a maximum at
certain intermediate value of $v$. This is exactly what we have seen
in Figs. \ref{fig:sound1D}, \ref{fig:sound2D} and \ref{fig:sound3D}
except in Fig.\ref{fig:sound3D}(d) where we see two local maxima.
%--------------------------------------------------------------------------------------
%--------------------------------------------------------------------------------------
%-----------------------------------------------------------------------------------------

To better understand the behavior of the sound speed as a function of
the lattice strength $v$, we have also compute numerically the
effective mass $m^*$ and the compressibility $\kappa$ and
the results are plotted in Figs.\ref{fig:CM1D},\ref{fig:CM2D},\&\ref{fig:CM3D}.
It is clear from these figures that the compressibility $\kappa$
has different dependence on $v$ in different dimensions.
This agrees with our analytical results in the last section.
However, we notice that the increasing rate of $\kappa^{-1}$ with $v$
is quite close in all dimensions.

The situation is different for the effective mass $m^*$. In the last
section, we have shown that the effective mass $m^*$ has the same
dependence on $v$ (see Eq. (\ref{mass})) in all dimensions. However,
it is true only in the limit of weak lattices. As seen in the right
columns of Figs.\ref{fig:CM1D},\ref{fig:CM2D}, \&\ref{fig:CM3D}, the
effective mass $m^*$, as a function of $v$, behaves very differently
in different dimensions. In particular, the increasing rate of $m^*$
with $v$ in one dimension is orders of magnitude larger than the
increasing rate in three dimensions. The two-dimensional case is
right in the between. Since the sound speed is the result of
competition of $m^*$ and $\kappa$, the relatively small increasing
rate of $m^*$ with $v$ allows the sound speed oscillates with $v$ in
3D.
%In all Figs.\ref{fig:CM1D},\ref{fig:CM2D}, \&\ref{fig:CM3D}, we
%always find the sound speed decreases with the increase of
%interactions $c$ for the fixed optical strength $v$, which can be
%understood based on Eqs.(\ref{mass},\ref{38},\ref{J}).

\section{Experiments}
The speed of sound of a BEC in an optical lattice may be measured
with the similar technique that was used in Ref.\cite{Andrews} to
measure the sound speed of a BEC in a trap. Some complication is
expected due to the periodic modulation of the BEC density. Another
possible method is to employ Bragg
spectroscopy\cite{Bragg,Excitation,Du} to the excitation spectrum.
The speed of sound can be extracted from the slope of the linear
part of the excitation spectrum.

In typical experiments to date, the relevant parameters are as
follows: for a BEC in three-dimensional optical
lattice\cite{Greiner}, the atom occupancy per lattice is of the
order of $\langle n_{i}\rangle \approx 1\sim 3$,
$n_{0}=1.3\sim3.9\times10^{19}m^{-3}$,
$k_{L}=2\pi/\lambda_{L}=7.37\times10^6m^{-1}$, and $a_{s}=5.4nm$;
for a BEC in quasi-two-dimensional optical lattice\cite{P2}, the
atom occupancy per lattice can reach $\langle n_{i}\rangle \approx
170$, $n_{0}=3.6\times10^{20}m^{-3}$,
$k_{L}=2\pi/\lambda_{L}=7.37\times10^6m^{-1}$, and $a_{s}=5.4nm$;
for a BEC in quasi-one dimensional optical lattice\cite{P1},
$\langle n_{i}\rangle \approx 1000$, $n_{0}=2.8\times10^{20}m^{-3}$,
$k_{L}=2\pi/\lambda_{L}=7.9\times10^6m^{-1}$, and $a_{s}=5.4nm$.
These parameters correspond to $c=0.08$ for 1D optical lattice,
$c=0.11$ for 2D optical lattice, and $c=0.004\sim0.012$ for 3D
optical lattice. The depth of optical lattice $V_{0}$ can be changed
from $0E_{R}$ to $32E_{R}$\cite{Greiner}; it means that our
 $v$ can be changed from $0$ and $2$. To our knowledge,
the highest atomic density without lattice is
$n_0=3\times10^{21}m^{-3}$ for sodium\cite{Inouye}. For this high
density, we have $c=0.22$ with $k_L=1.07\times
10^{7}m^{-1}$\cite{Choi}. However, this is rather idealistic. The
other possible way to increase $c$ is to tune the scattering length
$a_{s}$ with Feshbach resonance\cite{Fesh,FeshBloch}.

\section{Conclusions}

We have studied the speed of sound, compressibility, and effective
mass of a Bose-Einstein condensate in an optical lattice both
analytically and numerically.  Special attentions have been paid to
the effect of the depth of the optical lattice $v$, the interatomic
interaction $c$ and the dimensionality $D$ on the sound speed. Our
investigation shows that the sound speed depends strongly on the
strength of the lattice. In the one-dimensional case, the speed of
sound falls monotonically with increasing lattice strength. The
dependence becomes much richer in two and three dimensions. In the
two-dimensional case, when the interaction is weak, the sound speed
first increases then decreases as the lattice strength increases.
For the three-dimensional case, the sound speed can even oscillate
with the lattice strength. These rich behaviors can be understood
in terms of competition between compressibility and effective mass.
Our analytical results at the limit of weak lattices also offer an
interesting perspective to the understanding: they show the lattice
component perpendicular to the sound propagation decreases the sound
speed while the lattice components parallel to the propagation
increases the sound speed.

\section{Acknowledgements}
We thank X. Du and D. J. Heinzen for helpful discussion. This work
is supported by the ``BaiRen'' program of the Chinese Academy of
Sciences, the NSF of China (10504040), and the 973 project of China
(2005B724500). Z. D. Zhang and Z. X. Liang is supported by the NSF
of China (10674139).

\appendix
\section{Preliminary notations}

Suppose $f\left(\vec{r}\right)$ to be a periodic function with the
periodicity of $\vec{R}$, given by
\begin{equation}
f\left(\vec{r}\right)=f\left(\vec{r}+\vec{R}\right),
\end{equation}
with
\begin{equation}
\vec{R}=m_{1}\vec{a}_{1}+m_{2}\vec{a}_{2}+m_{3}\vec{a}_{3},
\end{equation}
where $\vec{r}$ is the position vector, $\vec{a}_{1}$,
$\vec{a}_{2}$, and $\vec{a}_{3}$ are any three vectors not all in
the same plane, and $m_{1}$, $m_{2}$, and $m_{3}$ ranges through all
integral values. Corresponding to $\vec{a}_{i}$'s, there exist a set
of reciprocal vectors $\vec{b}_{j}$'s such that
\begin{equation}
\vec{a}_{i}\cdot\vec{b}_{j}=2\pi\delta_{ij}.
\end{equation}
We can expand the periodic function $f\left(\vec{r}\right)$ as its
Fourier coefficients $\mathscr F_{\vec n}(f)$ as defined by,
\begin{equation}
f\left(\vec{r}\right)=\sum_{\vec n}\mathscr F_{\vec
n}\left(f\right)\exp\Big\{{i\vec{n}\cdot\vec r}\Big\},
\end{equation}
with
\begin{equation}
\mathscr{F}_{\vec{n}}\left(V\right) =\frac{1}{\Omega}\int_{cell}
d\vec{r} f\left( \vec{r} \right) \exp\Big\{ -i\vec{n} \cdot
\vec{r}\Big\}, \label{6}
\end{equation}
and
\begin{equation}
\vec{n}=n_{1}\vec{b}_{1}+n_{2}\vec{b}_{2}+n_{3}\vec{b} _{3}.
\end{equation}
where the $n_{j}$ are integers. In the integration, $\Omega$ is the
volume of the primitive cell and the integration is over a primitive
cell.

\section{Solutions of the Gross-Pitaevskii equation in the weak
potential limit}

The time-independent Gross-Pitaevskii (GP) equation in the
three-dimensional case can be written as
\begin{equation}\label{A1}
-\frac12\bigtriangledown^2\psi(\vec r)+c{|\psi(\vec r)|}^2\psi(\vec
r)+V_{ar}(\vec r)\psi(\vec r)=\mu\psi(\vec r),
\end{equation}
where $V_{ar}(\vec r)$ is the periodic potential with the
periodicity of $\vec{R}$,
\begin{equation}\label{A1-1}
V_{ar}\left(\vec r\right)=V_{ar}\left(\vec{r}+\vec{R}\right).
\end{equation}
The Bloch-wave solutions of the GP equation (\ref{A1}) reads
\begin{equation}\label{A2}
\psi(\vec r)=\phi_{\vec k}(\vec r)e^{i\vec k\cdot\vec r},
\end{equation}
where $\vec k$ is the Bloch wavenumber and $\phi_{\vec k}(\vec r)$
is a periodic function with the same periodicity of Eq.
(\ref{A1-1}). Substituting Eq. (\ref{A2}) into Eq. (\ref{A1}), we
have the following equation for each Bloch wave state $\phi_{\vec
k}(\vec r)$
\begin{eqnarray}\label{A3}
-\frac12{(\bigtriangledown+i\vec k)}^2\phi_{\vec k}(\vec r)&+&
c{|\phi_{\vec k}(\vec r)|}^2\phi_{\vec k}(\vec r)\nonumber\\
+V_{ar}(\vec r)\phi_{\vec k}(\vec
r)&=&\mu\left(\vec{k}\right)\phi_{\vec k}(\vec r).
\end{eqnarray}
The set of eigenvalues $\mu(\vec{k})$ then forms Bloch bands.

Besides the GP equation (\ref{A1}), the Bloch wave function is also
subject to the normalization condition given by
\begin{equation}
\frac{1}{\Omega}\int_{cell} d\vec{r} {|\phi|}^2=1,
\end{equation}
which is equivalent to
\begin{equation}\label{A4}
\mathscr F_{\vec0}({|\phi|}^2)=1.
\end{equation}
For convenience, we have dropped the suffix $\vec{k}$ and the
coordinate vector $\vec{r}$ in $\phi_{\vec{k}}\left(\vec{r}\right)$

Expanding $\phi$ in terms of the potential strength as
\begin{equation}
\phi=\phi^{(0)}+\phi^{(1)}+\phi^{(2)}+\cdots,
\end{equation}
we get the zeroth, first, and second order forms of Eq. (\ref{A4}),
respectively,
\begin{equation}\label{A5}
{\mathscr F_{\vec0}({|\phi|}^2)}^{(0)}=\sum_{\vec n}{|\mathscr
F_{\vec n}(\phi^{(0)})|}^2=1,
\end{equation}
\begin{eqnarray}\label{A6}
{\mathscr F_{\vec0}({|\phi|}^2)}^{(1)}&=&\sum_{\vec n}\left(\mathscr
F_{\vec n}(\phi^{(0)}){\mathscr F_{\vec n}^*
(\phi^{(1)})}\right.\nonumber\\
&&\left.+\mathscr F_{\vec{n}}(\phi^{(1)}){\mathscr F_{\vec
n}^*(\phi^{(0)})}\right)=0,
\end{eqnarray}
\begin{eqnarray}\label{A7}
{\mathscr F_{\vec0}({|\phi|}^2)}^{(2)}&=&\sum_{\vec n}\left(\mathscr
F_{\vec n}(\phi^{(0)}){\mathscr F_{\vec n}^*
(\phi^{(2)})}+{|\mathscr F_{\vec n}(\phi^{(1)})|}^2\right.\nonumber\\
&&\left.+\mathscr F_{\vec n}(\phi^{(2)}){\mathscr F_{\vec
n}^*(\phi^{(0)})}\right)=0.
\end{eqnarray}
There is still an arbitrary phase in the above wave functions, which
satisfy both the GP equation and the normalization condition.
Therefore, we may impose a third condition
\begin{equation}\label{A8}
\frac{1}{\Omega}\int d\vec{r}|\phi|^2 \in \mathbb R,
\end{equation}
So that the Bloch states can be uniquely determined.

Before solving the GP equation (\ref{A1}), we have to set forth
another two specifications. First, we are only concerned with Bloch
states at $\vec k=0$. In this case, we rewrite Eq. (\ref{A3}) as
follows
\begin{equation}\label{A9}
-\frac12\bigtriangledown^2\phi+c{|\phi|}^2\phi+V_{ar}(\vec
r)\phi=\mu\phi,
\end{equation}
where we dropped the suffix $\vec0$ and the coordinate vector $\vec
r$ in $\phi_{\vec 0}(\vec r)$ for convenience. Expanding $\phi$ and
$\mu$ in terms of the potential strength, we get the zeroth, first
and second order forms of Eq. (\ref{A9}), respectively,
\begin{equation}\label{A10}
-\frac12\bigtriangledown^2\phi^{(0)}+c{|\phi^{(0)}|}^2\phi^{(0)}=\mu^{(0)}\phi^{(0)},
\end{equation}
\begin{eqnarray}\label{A11}
-\frac12\nabla^2\phi^{(1)}+c\left(2{|\phi^{(0)}|}^2\phi^{(1)}+
{\phi^{(0)}}^2{\phi^{(1)}}^*\right)\nonumber\\
+V_{ar}(\vec r)\phi^{(0)}=\mu^{(0)}\phi^{(1)}+\mu^{(1)}\phi^{(0)},
\end{eqnarray}
\begin{eqnarray}\label{A12}
-\frac12\nabla^2\phi^{(2)}+c\left(2{|\phi^{(0)}|}^2\phi^{(2)}+
{\phi^{(0)}}^2{\phi^{(2)}}^*\right.\nonumber\\
\left.+2\phi^{(0)}{|\phi^{(1)}|}^2+{\phi^{(0)}}^*{\phi^{(1)}}^2\right)
+V_{ar}(\vec r)\phi^{(1)}\nonumber\\
=\mu^{(0)}\phi^{(2)}+\mu^{(1)}\phi^{(1)}+\mu^{(2)}\phi^{(0)}.
\end{eqnarray}
Second, we are only concerned with the cases in which the external
potential $V_{ar}(\vec r)$ is symmetric in each cell, or in other
words, $V_{ar}(\vec r)$ is an even function. Combining it with the
condition that $V_{ar}(\vec r)$ is a real function, we immediately
have
\begin{equation}
\mathscr F_{\vec n}(V)=\mathscr F_{-\vec n}(V)\in\mathbb R.
\end{equation}
In the following, we will solve the GP equation for obtaining the
normalized Bloch state at $\vec{k}=0$.

\subsection{The zeroth-order correction of the GP equation}

From Eq. (\ref{A10}), we get the zeroth-order wave function and
chemical potential, respectively,
\begin{equation}\label{A13}
\phi^{(0)}=1,\quad\mu^{(0)}=c,
\end{equation}
which automatically meet the normalization condition (\ref{A5}).

\subsection{The first-order correction of the GP equation}
Substituting Eq. (\ref{A13}) into Eq. (\ref{A6}), we have
\begin{equation}
\mathscr F_{\vec0}(\phi^{(1)})+{\mathscr F_{\vec0}^*(\phi^{(1)})}=0.
\end{equation}
From the phase condition (\ref{A8}), we know that $\mathscr
F_{\vec0}(\phi^{(1)})$ is a real number, and therefore
\begin{equation}\label{A14}
\mathscr F_{\vec0}(\phi^{(1)})=0.
\end{equation}
Substituting Eq. (\ref{A13}) into Eq. (\ref{A11}), we have
\begin{eqnarray}\label{A15}
\frac{1}{2}|\vec{n}|^2\mathscr F_{\vec
n}(\phi^{(1)})+c\left(\mathscr F_{\vec n}(\phi^{(1)})+\mathscr
F_{\vec n}({\phi^{(1)}}^*)
\right)\nonumber\\
+\mathscr F_{\vec n}(V)=\mu^{(1)}\delta_{\vec
n\vec0}.
\end{eqnarray}
Plugging Eq. (\ref{A14}) into Eq. (\ref{A15}) and letting $\vec
n=0$, we get the first-order correction of the chemical potential
\begin{equation}\label{A16}
\mu^{(1)}=\mathscr F_{\vec 0}(V).
\end{equation}
Taking complex conjugates on both sides of Eq. (\ref{A15}) and
replacing $-\vec n$ with $\vec n$, we obtain
\begin{eqnarray}\label{A17}
\frac{1}{2}|\vec{n}|^2\mathscr F_{\vec
n}({\phi^{(1)}}^*)+c\left(\mathscr F_{\vec n}(\phi^{(1)})+\mathscr
F_{\vec n}
({\phi^{(1)}}^*)\right)\nonumber\\
+\mathscr F_{\vec
n}(V)=\mu^{(1)}\delta_{\vec n\vec0}.
\end{eqnarray}
The unique solution of Eqs. (\ref{A15}) and (\ref{A17}) in the case
of $\vec n\ne0$ reads
\begin{equation}\label{A18}
\mathscr F_{\vec n}(\phi^{(1)})=\mathscr F_{\vec
n}({\phi^{(1)}}^*)=-\frac{\mathscr F_{\vec n}(V)}{\frac{1}{2}|\vec
n|^2+2c}, \quad\vec n\ne\vec0.
\end{equation}
From Eqs. (\ref{A14}) and (\ref{A18}), we know
\begin{equation}
\mathscr F_{\vec n}(\phi^{(1)})=\mathscr F_{-\vec
n}(\phi^{(1)})\in\mathbb R,
\end{equation}
which means that $\phi^{(1)}(\vec r)$ is a real even function.

\subsection{The second-order correction of the GP equation}
Plugging Eqs. (\ref{A13}) and (\ref{A14}) into Eq. (\ref{A7}), we
obtain
\begin{equation}\label{A19}
\mathscr F_{\vec0}(\phi^{(2)})+{\mathscr
F_{\vec0}(\phi^{(2)})}^*+\sum_{\vec n}{\mathscr F_{\vec
n}(\phi^{(1)})}^2=0.
\end{equation}
Plugging Eqs. (\ref{A13}) and (\ref{A14}) into Eq. (\ref{A12}), we
obtain
\begin{eqnarray}
&&\frac{1}{2}|\vec{n}|^2\mathscr F_{\vec
n}(\phi^{(2)})+c\left(\mathscr F_{\vec n}(\phi^{(2)})+\mathscr
F_{\vec n}({\phi^{(2)}}^*)
+3\mathscr F_{\vec n}({\phi^{(1)}}^2)\right)\nonumber\\
&&+\mathscr F_{\vec n}(V_{ar}\phi^{(1)})=\mu^{(1)}\mathscr F_{\vec
n}(\phi^{(1)}) +\mu^{(2)}\delta_{\vec{n}\vec{0}}.
\end{eqnarray}
In the case of $\vec n=0$, we have
\begin{eqnarray}\label{A20}
&&c\left(\mathscr F_{\vec0}(\phi^{(2)})+\mathscr
F_{\vec0}({\phi^{(2)}}^*)+
3\mathscr F_{\vec0}({\phi^{(1)}}^2)\right)+\mathscr F_{\vec0}(V\phi^{(1)})\nonumber\\
&&=\mu^{(1)}\mathscr F_{\vec0}(\phi^{(1)}) +\mu^{(2)}\delta_{\vec
n\vec0}.
\end{eqnarray}
Plugging Eqs. (\ref{A16}), (\ref{A18}) and (\ref{A19}) into Eq.
(\ref{A20}), we obtain the second-order correction of the chemical
potential
\begin{equation}\label{SecChem}
\mu^{(2)}=-\sum_{\vec{n}\ne\vec0}\frac{\frac{1}{2}|\vec
n|^2}{{\left(\frac{1}{2}|\vec n|^2+2c\right)}^2}{\mathscr F_{\vec
n}(V)}^2.
\end{equation}

To complete the calculation of the sound speed, we also need to
calculate the system energy near $\vec{k}=0$. This can be obtained
in terms of the effective potential
$c|\phi|^2+V_{ar}\left(\vec{r}\right)-\mu$ seen by each atom. We
view our system as a noninteracting gas in the effective potential,
\begin{equation}\label{Eff}
V_{eff}\left(\vec{r}\right)=\frac{|\vec{n}|^2}{|\vec{n}|^2+4c}V(\vec{r}).
\end{equation}
Since the correction to the system energy is second order in the
potential strength, it is sufficient to consider the first-order
correction of the Bloch state, there is no need of calculating the
second-order correction of the Bloch state. Based on Eq.
(\ref{Eff}), we can easily obtain the system energy
$E\left(\vec{k}\right)$ near $\vec{k}=0$, up to the second-order
correction,
\begin{eqnarray}
E\left(\vec{k}\right) =\frac{|\vec{k}| ^{2}}{2}-\sum_{ \vec{n}\neq
0}\frac{\frac{\left\vert \vec{n}\right\vert ^{4}}{\left( \left\vert
\vec{n}\right\vert ^{2}+4c\right) ^{2}}}{\frac{1}{2}\left( \vec{n}+
\vec{k}\right) ^{2}-\frac{1}{2}\left\vert \vec{k}\right\vert
^{2}}\mathscr{F} _{\vec{n}}^{2}\left( V\right).   \label{EffE}
\end{eqnarray}

\section{Analytical Expression of Sound Speed Based on Eq. (17) in Weak Potential Limit}

The aim of this section is to calculate the compressibility $\kappa$
and the effective mass $m^{*}$ as a function of the interatomic
interaction $c$ and of the depth of the arbitrary periodic potential
$V_{ar}\left(\vec{r}\right)$. Using these quantities, we will
calculate the velocity of sound.

\subsection{Compressibility $\kappa$ and effective mass $m^{*}$}
Plugging Eqs. (\ref{A13}), (\ref{A16}), and (\ref{SecChem}) into Eq.
(\ref{Comp}), we obtain the analytical expression of compressibility
$\kappa$ in the weak potential limit,
\begin{equation}\label{C1}
\kappa ^{-1}=c\left( 1-\sum_{\vec{n}\neq 0}\frac{16\left\vert
\vec{n} \right\vert ^{2}}{\left( \left\vert \vec{n}\right\vert
^{2}+4c\right) ^{3}} \mathscr{F}_{\vec{n}}^{2}\left( V\right)
\right).
\end{equation}

To calculate the sound speed, we also need calculating the effective
mass $m^{*}$. Substituting Eq. (\ref{EffE}) into Eq. (\ref{Emass}),
we obtain the analytical expression of effective mass along a given
direction indicated by a unit vector $\hat{r}$,
\begin{equation}\label{C1-1}
\frac{1}{m^{*}}=1-\sum_{\vec{n}\neq
0}\frac{16\left|\vec{n}\cdot\hat{r}\right|^{2}}{\left\vert \vec{n}
\right\vert ^{2}\left( \left\vert \vec{n}\right\vert ^{2}+4c\right)
^{2}} \mathscr{F}_{\vec{n}}^{2}\left( V\right).
\end{equation}

We also find that the effective mass along each axis $\vec{x}$,
$\vec{y}$, and $\vec{z}$ labeled by $m^{*}_{x}$, $m^{*}_{y}$, and
$m^{*}_{z}$ read,
\begin{equation}\label{C2}
\frac{1}{m^{*}_{x}}=1-\sum_{\vec{n}\neq
0}\frac{16\left|\vec{n}\cdot\vec{x}\right|^{2}}{\left\vert \vec{n}
\right\vert ^{2}\left( \left\vert \vec{n}\right\vert ^{2}+4c\right)
^{2}} \mathscr{F}_{\vec{n}}^{2}\left( V\right),
\end{equation}
and
\begin{equation}\label{C3}
\frac{1}{m^{*}_{y}}=1-\sum_{\vec{n}\neq
0}\frac{16\left|\vec{n}\cdot\vec{y}\right|^{2}}{\left\vert \vec{n}
\right\vert ^{2}\left( \left\vert \vec{n}\right\vert ^{2}+4c\right)
^{2}} \mathscr{F}_{\vec{n}}^{2}\left( V\right),
\end{equation}
and
\begin{equation}\label{C4}
\frac{1}{m^{*}_{z}}=1-\sum_{\vec{n}\neq
0}\frac{16\left|\vec{n}\cdot\vec{z}\right|^{2}}{\left\vert \vec{n}
\right\vert ^{2}\left( \left\vert \vec{n}\right\vert ^{2}+4c\right)
^{2}} \mathscr{F}_{\vec{n}}^{2}\left( V\right).
\end{equation}

Plugging Eqs. (\ref{C1-1}) and (\ref{C1}) into Eq. (\ref{sound}), we
arrive at the analytical expression of sound speed labeled by
$v_{s}$ along a given direction $\hat{r}$,
\begin{eqnarray}
v_{s}=\sqrt{c}&+&8\sqrt{c}\Big\{\sum_{\vec{n}\ne0}\frac{|\vec{n}|^2}
{(4c+|\vec{n}|^2)^{3}}\nonumber\\
&-&\sum_{\vec{n}\ne0}\frac{\left|\vec{n}\cdot\hat{r}\right|^{2}}{|\vec{n}|^2(4c+|\vec{n}|^2)^{2}}
\Big\}{\mathscr F_{\vec n}}^2(V).\label{C4-1}
\end{eqnarray}

Plugging Eqs. (\ref{C2}), (\ref{C3}), (\ref{C4}), and (\ref{C1})
into Eq. (\ref{sound}), we  also obtain the analytical expressions
of sound speed along each axis $\vec{x}$, $\vec{y}$, and $\vec{z}$,
labeled by $v_{sx}$, $v_{sy}$, and $v_{sz}$,
\begin{eqnarray}\label{C5}
v_{sx}=\sqrt{c}&+&8\sqrt{c}\Big\{\sum_{\vec{n}\ne0}\frac{|\vec{n}|^2}
{\left(4c+|\vec{n}|^2\right)^{3}}\nonumber\\
&-&\sum_{\vec{n}\ne0}\frac{\left|\vec{n}\cdot\vec{x}\right|^{2}}
{|\vec{n}|^2\left(4c+|\vec{n}|^2\right)^{2}} \Big\}{\mathscr F_{\vec
n}}^2(V),
\end{eqnarray}
and
\begin{eqnarray}\label{C6}
v_{sy}=\sqrt{c}&+&8\sqrt{c}\Big\{\sum_{\vec{n}\ne0}\frac{|\vec{n}|^2}
{(4c+|\vec{n}|^2)^{3}}\nonumber\\
&-&\sum_{\vec{n}\ne0}\frac{\left|\vec{n}\cdot\vec{y}\right|^{2}}{|\vec{n}|^2(4c+|\vec{n}|^2)^{2}}
\Big\}{\mathscr F_{\vec n}}^2(V),
\end{eqnarray}
and
\begin{eqnarray}\label{C7}
v_{sz}=\sqrt{c}&+&8\sqrt{c}\Big\{\sum_{\vec{n}\ne0}\frac{|\vec{n}|^2}
{(4c+|\vec{n}|^2)^{3}}\nonumber\\
&-&\sum_{\vec{n}\ne0}\frac{\left|\vec{n}\cdot\vec{z}\right|^{2}}{|\vec{n}|^2(4c+|\vec{n}|^2)^{2}}
\Big\}{\mathscr F_{\vec n}}^2(V).
\end{eqnarray}

In the following, we consider a special case, i.e. $\vec{a}_{1}$,
$\vec{a}_{2}$, and $\vec{a}_{3}$ are chosen along $\vec{x}$,
$\vec{y}$, and $\vec{z}$ respectively; we also suppose
$|\vec{a}_{1}|=|\vec{a}_{2}|=|\vec{a}_{3}|=2\pi$, without loss of
generality. In this case, the sound speed of Eqs. (\ref{C5}),
(\ref{C6}), and (\ref{C7}) can be simplified into,
\begin{eqnarray}\label{C11}
v_{sx}=\sqrt{c}&+&8\sqrt{c}\Big\{\sum_{\vec{n}\ne0}\frac{\vec{n}^2}
{(4c+\vec{n}^2)^{3}}\nonumber\\
&-&\sum_{\vec{n}\ne0}\frac{n_{1}^{2}} {\vec{n}^2(4c+\vec{n}^2)^{2}}
\Big\}{\mathscr F_{\vec n}}^2(V),
\end{eqnarray}
and
\begin{eqnarray}\label{C12}
v_{sy}=\sqrt{c}&+&8\sqrt{c}\Big\{\sum_{\vec{n}\ne0}\frac{\vec{n}^2}
{(4c+\vec{n}^2)^{3}}\nonumber\\
&-&\sum_{\vec{n}\ne0}\frac{n_{2}^{2}}{\vec{n}^2(4c+\vec{n}^2)^{2}}
\Big\}{\mathscr F_{\vec n}}^2(V),
\end{eqnarray}
\begin{eqnarray}\label{C13}
v_{sz}=\sqrt{c}&+&8\sqrt{c}\Big\{\sum_{\vec{n}\ne0}\frac{\vec{n}^2}
{(4c+\vec{n}^2)^{3}}\nonumber\\
&-&\sum_{\vec{n}\ne0}\frac{n_{3}^{2}}{\vec{n}^2(4c+\vec{n}^2)^{2}}
\Big\}{\mathscr F_{\vec n}}^2(V).
\end{eqnarray}

\section{Analytical Expression of Sound Speed Based on Eq. (16) in Weak Potential Limit}
As shown in Section II.B, there are two equivalent ways to calculate
the velocity of sound; one is based on Eq. (16), the other comes
from Eq. (17). The aim of this section is to calculate the
analytical expression of sound speed based on Eq. (16) from the
another angle, by directly solving excitation energy $\epsilon(q)$.

\subsection{The matrices P, Q, S and T}
According to the Bogoliubov theory, the excitation energy
$\epsilon(q)$ of the BEC in the Bloch state at $\vec{k}=0$ can be
obtained by solving the following eigenvalue problem
\begin{equation}
\delta_{z}M\left(\vec{q}\right)
\begin{pmatrix}
u\cr v
\end{pmatrix}
=\epsilon (q)
\begin{pmatrix}
u\cr v
\end{pmatrix},
\end{equation}
with
\begin{equation}
M=\left(\begin{array}{cc} \mathscr L(\vec k+\vec q)&c{\phi_{\vec
k}}^2\\c{\phi_{\vec k}^*}^2&\mathscr L(-\vec k+\vec q)
\end{array}\right),
\end{equation}
and
\begin{equation}
\sigma_z=\left(\begin{array}{cc}1&0\\0&-1\end{array}\right),
\end{equation}
where $\mathscr L(\vec p)$ is defined as
\begin{equation}
\mathscr L(\vec q)=-\frac12{(\bigtriangledown+i\vec
q)}^2+V_{ar}(\vec r)-\mu+2c{|\phi_{\vec k}|}^2.
\end{equation}
By a similarity transformation, we can transform $\sigma_zM$ into a
numerical matrix $P$ without changing the eigenvalues. The new
matrix $P$ can be represented in a block form
\begin{equation}\label{A21}
P=(T_{\vec m\vec n})_{\infty\times\infty},
\end{equation}
where each block $T_{\vec m\vec n}$ is actually a $2\times2$ matrix
and $\vec m$ and $\vec n$ take values ranging from
$(-\infty,-\infty,-\infty)$ to $(+\infty,+\infty,+\infty)$. For
convenience, we abbreviate the diagonal blocks $T_{\vec n\vec n}$ in
Eq. (\ref {A21}) as $S_{\vec n}$, and consequently $T_{\vec m\vec
n}$ denotes solely those non-diagonal ($\vec m\ne\vec n$) blocks.

For we are only concerned with the case of $\vec k=0$, $\sigma_zM$
can be simplified into
\begin{equation}
\sigma_zM=\left(\begin{array}{cc}\mathscr L(\vec
q)&c\phi^2\\-c{\phi^*}^2&-\mathscr L(\vec q)\end{array}\right).
\end{equation}
In this case, we have
\begin{equation}\label{A22}
S_{\vec n}=\left(\begin{array}{cc}a_{\vec n}&b_{\vec n}\\-b_{\vec
n}^*&-a_{\vec n}\end{array}\right),\quad T_{\vec m\vec
n}=\left(\begin{array}{cc} c_{\vec m\vec n}&d_{\vec m\vec
n}\\-d_{\vec m\vec n}^*&-c_{\vec m\vec n}
\end{array}\right),
\end{equation}
with $a_{\vec n}$, $b_{\vec n}$, $c_{\vec m\vec n}$ and $d_{\vec
m\vec n}$ determined by
\begin{eqnarray}
&&a_{\vec n}=\frac12{(\vec n+\vec q)}^2+\mathscr F_{0}(V)-\mu+
2c\mathscr F_{0}({|\phi|}^2),\label{A23}\\
&&b_{\vec n}=c\mathscr F_{0}(\phi^2),\\
&&c_{\vec m\vec n}=\mathscr F_{\vec m-\vec n}(V)+2c\mathscr F_{\vec m-\vec n}({|\phi|}^2),\\
&&d_{\vec m\vec n}=c\mathscr F_{\vec m-\vec n}(\phi^2).\label{A24}
\end{eqnarray}
We define the matrix $Q$ as follows
\begin{equation}
Q=P-\varepsilon I,
\end{equation}
where $\varepsilon$ represents the lowest elementary excitation.
Consequently, we have
\begin{equation}\label{A25}
\det Q=0.
\end{equation}
Now, we compute the elements $a_{\vec n}$, $b_{\vec n}$, $c_{\vec
m\vec n}$ and $d_{\vec m\vec n}$ from Eqs. (\ref{A23}) --
(\ref{A24}) as follows
\begin{eqnarray}
a_{\vec n}^{(0)}&=&\frac12{(\vec n+\vec q)}^2+c, \ b_{\vec n}^{(0)}=c,\\
c_{\vec m\vec n}^{(0)}&=&0, \ d_{\vec m\vec n}^{(0)}=0,\label{A26}\\
a_{\vec n}^{(1)}&=&0, \ b_{\vec n}^{(1)}=0,\\
c_{\vec m\vec n}^{(1)}&=&\frac{\frac12{(\vec m-\vec n)}^2-2c}{\frac12{(\vec m-\vec n)}^2+2c}\mathscr F_{\vec m-\vec n}(V),\\
d_{\vec m\vec n}^{(1)}&=&-\frac{2c}{\frac12{(\vec m-\vec n)}^2+2c}\mathscr F_{\vec m-\vec n}(V),\\
a_{\vec n}^{(2)}&=&\sum_{\vec n\ne\vec0}\frac{\frac12\vec
n^2}{{\left(\frac12\vec n^2+2c\right)}^2}
{\mathscr F_{\vec n}(V)}^2,\\
b_{\vec n}^{(2)}&=&0.
\end{eqnarray}
For all these quantities are real, we may drop the asterisks * in
Eq. (\ref {A22}).

\subsection{The lowest elementary excitation in the weak potential limit}
We expand the matrix $Q$ and $\varepsilon$ in terms of $v$ as
follows
\begin{eqnarray}
Q&=&Q^{(0)}+Q^{(1)}+Q^{(2)}+\cdots,\\
\varepsilon&=&\varepsilon^{(0)}+\varepsilon^{(1)}+\varepsilon^{(2)}+\cdots.
\end{eqnarray}
The aim of this subsection is to calculate $\varepsilon^{(0)}$,
$\varepsilon^{(1)}$ and $\varepsilon^{(2)}$ by expanding Eq.
(\ref{A25}) into its zeroth, first and second order form.

\subsubsection{The zeroth-order approximation of $\varepsilon$}
The zeroth-order form of Eq. (\ref{A25}) is
\begin{equation}
{(\det Q)}^{(0)}=\det Q^{(0)}=\det(P^{(0)}-\varepsilon^{(0)}I)=0.
\end{equation}
From Eqs. (\ref{A26}), we know that all the $T_{\vec m\vec n}^{(0)}$
are zero matrices, and therefore the matrix $P^{(0)}$ is block
diagonal as represented in Eq. (\ref{A21}). Consequently, the
eigenvalues of $P^{(0)}$ are the collection of the eigenvalues of
each $S_{\vec n}^{(0)}$. The zeroth-order approximation of
$\varepsilon$ is hence the positive eigenvalue of $S_{0}^{(0)}$
\begin{equation}\label{A28}
\left|\begin{array}{cc}
a_{\vec0}^{(0)}-\varepsilon^{(0)}&b_{\vec0}^{(0)}\\-b_{\vec0}^{(0)}&-a_{\vec0}^{(0)}-\varepsilon^{(0)}
\end{array}\right|=0.
\end{equation}
The positive solution of Eq. (\ref{A28}) reads
\begin{equation}\label{Azero}
\varepsilon^{(0)}=\sqrt{\frac14\vec q^4+c\vec q^2}.
\end{equation}
With the value of $\varepsilon^{(0)}$, we calculate the determinant
of each diagonal block of the matrix $Q^{(0)}$
\begin{equation}
\det(S_{\vec n}^{(0)}-\varepsilon^{(0)}I)={\left(\frac12\vec
q^2+c\right)}^2-{\left(\frac12{(\vec n+\vec q)}^2 +c\right)}^2,
\end{equation}
which are denoted by $\mathscr R_{\vec n}$ for convenience
\begin{equation}
\mathscr R_{\vec n}={\left(\frac12\vec
q^2+c\right)}^2-{\left(\frac12{(\vec n+\vec q)}^2+c\right)}^2.
\end{equation}
This result will be useful in the following sections.

\subsubsection{The first-order correction of $\varepsilon$}
We can conclude that the first-order correction of $\varepsilon$
vanishes as
\begin{equation}\label{Afirst}
\varepsilon^{(1)}=0.
\end{equation}

\subsubsection{The second-order correction of $\varepsilon$}
The second-order form of Eq. (\ref{A25}) reads
\begin{eqnarray}\label{A32}
&&{(\det Q)}^{(2)}=\sum_{ij}{\left(\frac{\partial|Q|}{\partial Q_{ij}}\right)}^{(0)}Q_{ij}^{(2)}\nonumber\\
&&+\frac12\sum_{ijkl}{\left(\frac{\partial^2|Q|}{\partial
Q_{ij}\partial Q_{kl}}\right)}^{(0)}Q_{ij}^{(1)}Q_{kl}^{(1)}=0.
\end{eqnarray}
Here we introduce the ``second cofactor matrix'' of $Q$, which is
denoted by $\widetilde{\widetilde Q}$ and whose elements are defined
as
\begin{equation}
\widetilde{\widetilde Q}_{ij,kl}=\frac{\partial^2|Q|}{\partial
Q_{ij}\partial Q_{kl}}.
\end{equation}
By this notation, we reduce Eq. (\ref{A32}) into
\begin{equation}\label{A33}
\sum_{ij}\widetilde
Q_{ij}^{(0)}Q_{ij}^{(2)}+\frac12\sum_{ijkl}\widetilde{\widetilde
Q}_{ij,kl}^{(0)}Q_{ij}^{(1)} Q_{kl}^{(1)}=0.
\end{equation}

We start by computing the first term on the left hand side of Eq.
(\ref{A33}). Keeping only the non-vanishing terms and noting that
$\mathscr R_{\vec0}=0$, we have
\begin{eqnarray}\label{A34}
\sum_{ij}\widetilde Q_{ij}^{(0)}Q_{ij}^{(2)}&=&\sum_{\vec
n}(2\varepsilon^{(0)}\varepsilon^{(2)}-2a_{\vec n}^{(0)}
a_{\vec n}^{(2)})\prod_{\vec m\ne\vec n}\mathscr R_{\vec m}\nonumber\\
&=&(2\varepsilon^{(0)}\varepsilon^{(2)}-2a_{\vec0}^{(0)}a_{\vec0}^{(2)})\prod_{\vec
m\ne\vec0}\mathscr R_{\vec m}.
\end{eqnarray}

Then we proceed to compute the second term on the left hand side of
Eq. (\ref{A33}). It should be noted that any $Q_{ij}$ is one of the
elements of the matrix $U_{\vec m\vec n}$ which is defined as
\begin{equation}
U_{\vec m\vec n}=\left(\begin{array}{cccc}
a_{\vec m}-\varepsilon&b_{\vec m}&c_{\vec m\vec n}&d_{\vec m\vec n}\\
-b_{\vec m}&-a_{\vec m}-\varepsilon&-d_{\vec m\vec n}&-c_{\vec m\vec n}\\
c_{\vec n\vec m}&d_{\vec n\vec m}&a_{\vec n}-\varepsilon&b_{\vec n}\\
-d_{\vec n\vec m}&-c_{\vec n\vec m}&-b_{\vec n}&-a_{\vec
n}-\varepsilon
\end{array}\right).
\end{equation}

The second term on the left hand side of Eq. (\ref{A33}) could be
computed in a routine way by computing the second cofactor matrix of
$U_{\vec m\vec n}$. However, we have found a much more convenient
method to compute this term which is shown as follows. Since we have
\begin{equation}
a_{\vec m}^{(1)}-\varepsilon^{(1)}=-a_{\vec
m}^{(1)}-\varepsilon^{(1)}=b_{\vec m}^{(1)}=-b_{\vec m}^{(1)}=0,
\end{equation}
the only non-vanishing $Q_{ij}^{(1)}$ are those in $T_{\vec m\vec
n}^{(1)}$, or more specifically, $c_{\vec m\vec n}^{(1)}$, $-c_{\vec
m\vec n}^{(1)}$, $d_{\vec m\vec n}^{(1)}$ and $-d_{\vec m\vec
n}^{(1)}$. We further assert that in order to get a non-vanishing
$\widetilde{\widetilde Q}^{(0)}_{ij,kl}$, the corresponding $Q_{ij}$
and $Q_{kl}$ must be in the same $U_{\vec m\vec n}$. This assertion
can be confirmed by some routine proof which we shall not elaborate
here. From the above assertion, we obtain the following expression
\begin{eqnarray}\label{A35}
\sum_{ijkl}\widetilde{\widetilde
Q}^{(0)}_{ij,kl}Q_{ij}^{(1)}Q_{kl}^{(1)}=\nonumber\\
\sum_{\vec m,\vec n}^{\vec m\ne\vec n}
\sum_{ijkl=1}^4(\widetilde{\widetilde{U_{\vec m\vec
n}}})_{ij,kl}^{(0)}{(U_{\vec m\vec n})}_{ij}^{(1)} {(U_{\vec m\vec
n})}_{kl}^{(1)}\prod_{\vec k\ne\vec m,\vec n}\mathscr R_{\vec k}.
\end{eqnarray}
Keeping only the second-order correction terms, we have
\begin{eqnarray}\label{A36}
&&\sum_{ijkl=1}^4(\widetilde{\widetilde{U_{\vec m\vec
n}}})_{ij,kl}^{(0)}{(U_{\vec m\vec n})}_{ij}^{(1)}
{(U_{\vec m\vec n})}_{kl}^{(1)}\nonumber\\
&&=-2{\varepsilon^{(0)}}^2c_{\vec m\vec n}^{(1)}c_{\vec n\vec
m}^{(1)}-2a_{\vec m}^{(0)}a_{\vec n}^{(0)} c_{\vec m\vec
n}^{(1)}c_{\vec n\vec m}^{(1)}-2b_{\vec m}^{(0)}b_{\vec
n}^{(0)}c_{\vec m\vec n}^{(1)}
c_{\vec n\vec m}^{(1)}\nonumber\\
&&+2a_{\vec n}^{(0)}b_{\vec m}^{(0)}c_{\vec n\vec m}^{(1)}d_{\vec
m\vec n}^{(1)}+2a_{\vec m}^{(0)}b_{\vec n}^{(0)}
c_{\vec n\vec m}^{(1)}d_{\vec m\vec n}^{(1)}\nonumber\\
&&+2a_{\vec n}^{(0)}b_{\vec m}^{(0)}c_{\vec m\vec n}^{(1)}d_{\vec
n\vec m}^{(1)}+2a_{\vec m}^{(0)}b_{\vec n}^{(0)}
c_{\vec m\vec n}^{(1)}d_{\vec n\vec m}^{(1)}+2{\varepsilon^{(0)}}^2d_{\vec m\vec n}^{(1)}d_{\vec n\vec m}^{(1)}\nonumber\\
&&-2a_{\vec m}^{(0)}a_{\vec n}^{(0)}d_{\vec m\vec n}^{(1)}d_{\vec
n\vec m}^{(1)}-2b_{\vec m}^{(0)}b_{\vec n}^{(0)} d_{\vec m\vec
n}^{(1)}d_{\vec n\vec m}^{(1)},
\end{eqnarray}
which is denoted by $\mathscr W_{\vec m\vec n}$ for convenience.
Since both $c_{\vec m\vec n}^{(1)}$ and $d_{\vec m\vec n}^{(1)}$ are
symmetric in $\vec m$ and $\vec n$, the same is true for $\mathscr
W_{\vec m\vec n}$
\begin{equation}
\mathscr W_{\vec m\vec n}=\mathscr W_{\vec n\vec m}.
\end{equation}

Plugging Eq. (\ref{A36}) into Eq. (\ref{A35}) and noting that
$\mathscr R_{\vec0}=0$, we obtain
\begin{eqnarray}\label{A37}
\sum_{ijkl}\widetilde{\widetilde
Q}^{(0)}_{ij,kl}Q_{ij}^{(1)}Q_{kl}^{(1)}=2\sum_{\vec
n\ne\vec0}\frac{\mathscr W_{\vec n\vec0}}{\mathscr R_{\vec
n}}\prod_{\vec m\ne\vec0}\mathscr R_{\vec n}.
\end{eqnarray}
Plugging Eqs. (\ref{A34}) and (\ref{A37}) into Eq. (\ref{A33}), we
have
\begin{equation}
(2\varepsilon^{(0)}\varepsilon^{(2)}-2a_{\vec0}^{(0)}a_{\vec0}^{(2)})\prod_{\vec
m\ne\vec0}\mathscr R_{\vec m} +\sum_{\vec n\ne\vec0}\frac{\mathscr
W_{\vec n\vec0}}{\mathscr R_{\vec n}}\prod_{\vec m\ne\vec0}\mathscr
R_{\vec m}=0,
\end{equation}
from which we finally arrive at the value of $\varepsilon^{(2)}$
\begin{equation}\label{Asecond}
\varepsilon^{(2)}=\frac1{\varepsilon^{(0)}}\sum_{\vec
n\ne\vec0}\left(a_{\vec0}^{(0)}\frac{\frac{1}{2}|\vec
n|^2}{{\left(\frac{1}{2}|\vec{n}|^2 +2c\right)}^2}{\mathscr F_{\vec
n}(V)}^2-\frac12\frac{\mathscr W_{\vec n\vec0}}{\mathscr R_{\vec
n}}\right).
\end{equation}

\subsection{The speed of sound in the weak potential limit}
The speed of sound along any given direction $\hat r$ in a BEC is
defined as
\begin{equation}
v_{\hat n}=\left|\nabla_{\vec q}\,\varepsilon\left(\vec
q\right)\right|_{\vec q\to\vec0^+}.
\end{equation}
Let's consider a special case, i.e. $\vec{a}_{1}$, $\vec{a}_{2}$,
and $\vec{a}_{3}$ are chosen along each axis of reference system,
$\vec{x}$, $\vec{y}$, and $\vec{z}$, respectively; and without loss
of generality, we suppose the periodicity of the periodic potential
along each axis to be $2\pi$. We are particularly interested in the
speed of sound along one of the three axes, for example x-axis
\begin{equation}\label{Asound}
v_s={\left.\frac{\partial\varepsilon}{\partial q_x}\right|}_{\vec
q\to0^+}.
\end{equation}
Plugging Eqs. (\ref{Azero}), (\ref{Afirst}) and (\ref{Asecond}) into
Eq. (\ref{Asound}), we finally obtain the analytical expression of
the sound speed along each axis,
\begin{equation}\label{A99}
v_{sx}=\sqrt{c}+\sum_{\vec{n}\ne0}\frac{8\sqrt{c}
\left[(n_2^2+n_3^2)|\vec{n}|^2-4cn_1^2\right]}
{|\vec{n}|^2\left(4c+|\vec{n}|^2\right)^3}{\mathscr F_{\vec
n}}^2(V),
\end{equation}
and
\begin{equation}\label{A99-1}
v_{sy}=\sqrt{c}+\sum_{\vec{n}\ne0}\frac{8\sqrt{c}
\left[(n_1^2+n_3^2)|\vec{n}|^2-4cn_2^2\right]}
{|\vec{n}|^2\left(4c+|\vec{n}|^2\right)^3}{\mathscr F_{\vec
n}}^2(V),
\end{equation}
and
\begin{equation}\label{A99-2}
v_{sz}=\sqrt{c}+\sum_{\vec{n}\ne0}\frac{8\sqrt{c}
\left[(n_1^2+n_2^2)|\vec{n}|^2-4cn_3^2\right]}
{|\vec{n}|^2\left(4c+|\vec{n}|^2\right)^3}{\mathscr F_{\vec
n}}^2(V).
\end{equation}
It can be easily proved that Eqs. (\ref{A99})-(\ref{A99-2}) can be
deduced into Eqs. (\ref{C11})-(\ref{C13}).

\section{A Special Example}

Suppose that the arbitrary potential of $V_{ar}(\vec{r})$ is chosen
to the special form of Eq. (\ref{pot})
\begin{equation}
V_{latt}(\vec{r})=v\left(\cos x+\cos y+\cos z\right).
\end{equation}
In this case, there are only six non-vanishing Fourier coefficients
\begin{equation}\label{A102}
\mathscr F_{\vec n}(V)=\frac{v}{2}\left(\delta_{\vec{n},(\pm1,0,0)}+
\delta_{\vec{n},(0,\pm1,0)}+\delta_{\vec{n},(0,0,\pm1)}\right).
\end{equation}
Submitting Eq. (\ref{A102}) into Eq. (\ref{A99}), we have
\begin{equation}\label{A103}
v_{s}=\sqrt{c}\left(1+\frac{8\left(1-2c\right)}{\left(4c+1\right)^3}v^2\right),
for D=3.
\end{equation}
With the similar calculations, we can also obtain the analytic
expressions of sound speed in the optical lattice of Eqs. (7) and
(8), respectively,
\begin{eqnarray}
v_{sx}&=&\sqrt{c}\left(1+\frac{16c}{\left(4c+1\right)^3}v^2\right),
for D=1,\\ \label{A104}
v_{s}&=&\sqrt{c}\left(1+\frac{8\left(1-4c\right)}{\left(4c+1\right)^3}v^2\right),
for D=2.\label{A105}
\end{eqnarray}
Combining Eqs. (\ref{A103}), (\ref{A104}) and (\ref{A105}) together,
we arrive
\begin{equation}\label{E6}
v_{s}= \sqrt{c}\left( 1+\frac{4(D-1-4c)}{\left(4c+1\right)
^{3}}v^{2}\right).
\end{equation}

\end{document}